\documentclass[prd,aps,preprint,showpacs,nofootinbib,superscriptaddress]{revtex4}
\usepackage{epsfig}
\usepackage{amsmath}

\begin{document}

\preprint{SLAC-PUB-13874}

\title{Phases of Augmented Hadronic Light-Front Wave Functions}

\author{Stanley J. Brodsky}
\affiliation{SLAC National Accelerator Laboratory, Stanford
University, Stanford, CA 94309}
\author{Barbara Pasquini}
\affiliation{Dipartimento di Fisica Nucleare e Teorica, Universit\`{a} degli Studi di Pavia,
and INFN, Sezione di Pavia, Italy }
\author{Bo-Wen Xiao}
\affiliation{Nuclear Science Division,
Lawrence Berkeley National Laboratory, Berkeley, CA 94720}
\author{Feng Yuan}
\affiliation{Nuclear Science Division,
Lawrence Berkeley National Laboratory, Berkeley, CA
94720}\affiliation{RIKEN BNL Research Center, Building 510A,
Brookhaven National Laboratory, Upton, NY 11973}

\begin{abstract}
It is an important question whether the final/initial state gluonic
interactions which lead to naive-time-reversal-odd single-spin asymmetries  and
diffraction at leading twist can be associated in a definite way
with the light-front wave function hadronic eigensolutions of QCD.
We use  light-front time-ordered perturbation theory to obtain
augmented light-front wave functions  which contain an imaginary
phase which depends on the choice of advanced or retarded
boundary condition for the gauge potential in light-cone gauge.
We apply this formalism to the wave functions of  the valence
Fock states of nucleons and pions, and show how this illuminates
the factorization properties of naive-time-reversal-odd transverse
momentum dependent observables which arise from rescattering.
In particular, one calculates the identical leading-twist Sivers function
from the overlap of augmented light-front wavefunctions that one obtains from
explicit calculations of the single-spin asymmetry in semi-inclusive deep inelastic lepton-polarized nucleon scattering where the required phases come from the final-state rescattering of the struck quark  with the nucleon spectators.
\end{abstract}

\maketitle

\section{Introduction}

Wave functions are key objects of the quantum world, specifying the
structure of composite states in terms of their fundamental constituents.
The conceptual extension of the  non-relativistic wavefunctions of
Schr\"odinger theory  to relativistic hadron physics are the frame-independent
light-front wave functions (LFWFs) of
hadrons $\Psi^H_n(x_i, \vec k_{i\perp}, \lambda_i)$
where $x_i  = {k^+_i\over P^+} = { k^0_i+k^z_i\over P^0 + P^z}$
are the light-front momentum fractions of the $n$ constituents,
$k_{i\perp}$ the transverse momentum components, and $\lambda_i$
the parton helicities.
The LFWFs are defined as  constituent wave functions at fixed
light-front time $\tau = x^+ = t +  z/c$
and in the light-cone gauge $A^+=A^0+A^z=0$ where
$A^\mu$ represents the gauge field~\cite{bl,Lepage:1980fj,
Brodsky:1997de,Burkardt:1995ct}.  The LFWFs are obtained
explicitly by computing the hadronic eigensolutions
$|\Psi_H \rangle$ of the QCD light-front Hamiltonian $H_{LF}$
projected on the free Fock basis $\Psi^H_n= \langle n|\Psi_H\rangle.$

Light-front wave functions in QCD describe the quark and
gluon composition of hadron at a fundamental level,
leading to a description of a wide range of
hadronic and nuclear physics phenomena~\cite{Lepage:1980fj}.
For example, the parton distribution functions measured
in deep inelastic lepton-hadron scattering, including
DGLAP evolution and their transverse momentum extensions, are defined from
the sum over squares of the light-front wave functions.
Form factors are given by the sum of overlap matrix
elements of the initial and final LFWFs with the electroweak currents.
The gauge-invariant distribution amplitudes $\phi(x_i, Q)$
which control hard exclusive reactions are the valence LFWFs
integrated over transverse momenta $k^2_\perp < Q^2.$

Recent theoretical developments
have shown that final and/or initial-state interactions
can generate a phase in scattering amplitudes
which lead to novel single transverse spin asymmetries
in high energy hadronic reactions at leading twist.  A prime
example of this rescattering physics in QCD is the Sivers
single-spin asymmetry  measured in semi-inclusive deep inelastic
scattering and spin-dependent Drell-Yan
lepton pair production~\cite{BroHwaSch02,Col02,{BelJiYua02}}.
Double initial-state interactions lead to an anomalous
$\cos 2 \phi$ azimuthal
dependence of the production plane in unpolarized
lepton pair hadroproduction, corresponding  to
the breakdown of the Lam-Tung relation
in PQCD~\cite{Boer:2002ju,{Zhou:2009rp}}.
Similarly, diffractive deep inelastic lepton scattering
$\ell p \to \ell^\prime p^\prime X$ arises from the exchange
of gluons in the final state which occurs after the hard
lepton-quark interaction~\cite{Brodsky:2002ue}.  Since
nuclear shadowing involves diffractive deep inelastic
processes, nuclear distributions are also dependent
on rescattering processes.

The wavefunctions of stable hadrons that are obtained
by solving the Heisenberg problem
$H^{QCD}_{LF} |\Psi> = {\cal M}^2 |\Psi>$  have a real phase.
As discussed in Ref.~\cite{Brodsky:2009dv}, one can
distinguish ``static" structure functions, the
probabilistic distributions computed from the square of
the light-front wavefunctions of the target hadron  from
the ``dynamic" structure functions measured in deep
inelastic lepton-hadron scattering which include the
effects of rescattering associated with the Wilson line.
Thus it is an important question whether the final/initial
state gluonic interactions responsible for  the dynamics of
rescattering can be associated in a definite way with
the light-front wave function eigensolutions of QCD.
The resulting augmented LFWFs provide an
important tool for understanding the factorization properties of
dynamical hadronic phenomena including single-spin asymmetries and diffraction.

It has been shown that the light-cone gauge condition $A^+=0$ does not fix
the gauge of Abelian or non-Abelian  gauge fields completely~\cite{BelJiYua02}:
one has to choose a boundary
condition for the transverse component of the gauge
potential at spatial infinity: $A_{\perp}(x^-=\pm \infty)$~\cite{BelJiYua02}.
The propagators of the gauge field which define  the QCD Light-Front
Hamiltonian in the Heisenberg problem are regulated using
the principal value prescription.
However, a different choice of boundary condition
will lead to different properties of the
light-front wave function amplitudes. In particular, if we
choose a retarded ($A_\perp(x^-=-\infty)=0$)
or advanced ($A_\perp(x^-=\infty)=0$) boundary
condition, the resulting augmented light-front wave function will contain
the necessary phase to generate the nonzero
single spin asymmetry in hadronic reactions.
We will demonstrate these properties,  giving
an explicit calculation in light-front time-ordered
perturbation theory~\cite{Lepage:1980fj}.
The result  of our analysis provides the general structure of augmented LFWFs
which is easy to apply to phenomenological applications.
As an example, we will present results for the
three-quark Fock state component of nucleon and the quark-antiquark
component of pion at lowest non-trivial order. We can further simplify
the result for the pion in terms of the distribution amplitudes.
Given these light-front wave function amplitudes results,
it is straightforward to calculate the pseudo-time-reversal-odd
quark distributions of the nucleon and pion, by applying
the overlap formalism derived in~\cite{Ji:2002xn}.
The light-cone gauge with retarded/advanced boundary condition
has also been used to investigate the small-$x$ physics~\cite{Mueller:1985wy,Kovchegov:1997pc,mueller,raju},
in particular, to study the evolution and
factorization for nucleus-nucleus collisions~\cite{raju}.

The rest of this paper is organized as follows. In Sec. II,
we present a general derivation of augmented LFWFs
using light-front time-order
perturbation theory within a lowest order formalism.
In Sec. III, we apply our method to the construction of
augmented light-front wave function amplitudes
for the three-quark Fock component of nucleon and quark-antiquark
component of the pion. We summarize our paper in Sec. IV.

\section{General Derivations}
We start our derivation by constructing the general form for a Fock state
expansion of any given hadron,
\begin{eqnarray}
|P,S\rangle= \sum_n     \int \prod_{i=1}^n d[i]
       \psi_n(x_i,k_{i\perp},\lambda_i)~
    a_1^\dagger a_2^\dagger ... a_n^\dagger  |0\rangle \ ,
\end{eqnarray}
where $P$ and $S$ are the momentum and spin of the hadron,
$d[i] = dx_id^2k_{i\perp}/(\sqrt{2x_i}(2\pi)^3)$ with the
overall constraint on $x_i$ and $k_{i\perp}$ implicit.
For convenience, in following calculations
we set the transverse momentum of hadron equal
zero: $P_\perp=0$. Because the wave functions
are boost invariant, all our results can be extended to more
general case with $P_\perp\neq 0$. In this
Fock state, each parton is represented by the associated
creation operator $a_i^\dagger(k_i)$ with $k^\mu_i = (k^+_i, \vec k_{\perp i})
= (x_iP^+, \vec k_{\perp i})$,
which contains certain longitudinal momentum
$k_i^+=x_iP^+$ and transverse momentum $k_{i\perp}$,
whereas the minus component is determined by the on-shell condition
$k_i^-=(k_\perp^2+m_i^2)/k_i^+$.
Implicitly, the above light-front wave function amplitude $\psi_n$
depends on the orbital angular momentum
projection from the constituents with the form of
($k_{i}^x\pm k_i^y$)~\cite{Ji:2003fw}.
Since the following derivation does not depend on this
structure, we will not include it explicitly.

As discussed in the Introduction, the definition of
a  light-front wave function amplitude $\psi_n$ can
be extended to include rescattering effects so its
phase is not necessary real. We can obtain the imaginary
part (or the phase), by iterating the light-front
wave function eigensolutions employing a particular
boundary condition for the gauge field.  In fact as
we shall show, the  phase of the augmented wavefunctions
can be computed perturbatively by applying
light-front time-ordered perturbation theory~\cite{Lepage:1980fj}
analogous to the Lippmann-Schwinger method.

The first order correction to the LFWF can be obtained
by iterating the Light-Front equation of motion:
\begin{eqnarray}
(P^--\small{\sum} k^-)\psi_n(x_i,k_{i\perp})=\int d[i]' K[k;\ell]\otimes \psi_n'(y_i,\ell_{i\perp}) \ ,
\end{eqnarray}
where $\sum k^-$ represents the sum of all partons energy $k_i^-$,
$d[i]'$ represents the integral of $(y_i,\ell_{i\perp})$. The interaction
kernel $K$ can be calculated from the light-front time-order perturbation
theory~\cite{Lepage:1980fj}. The wave functions $\psi_n$ and $\psi'_n$ may differ.
From the above expression, we find that the phase of $\psi_n$ may come
from the wave function in the right hand side $\psi'_n$ or the
interaction kernel $K$. In the following, we assume that the wave
function $\psi'_n$ is real, for example, from model calculation such
as constituent quark model~\cite{barbara}. We will focus on the contribution from
the interaction kernel. We will calculate, in particular,
the one-gluon exchange contribution to the interaction kernel.

\begin{figure}[t]
\begin{center}
\includegraphics[height=4.0cm,angle=0]{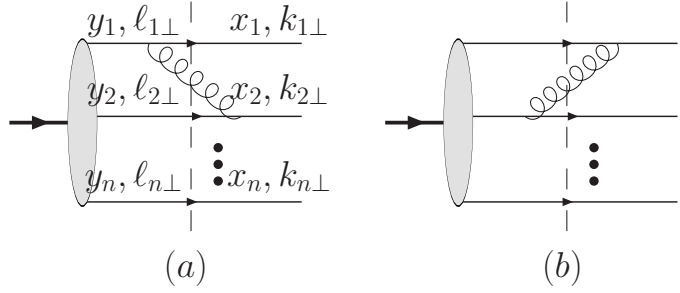}
\vspace*{-0.2cm} \epsfysize=4.2in
\end{center}
\caption{\it Light-front time-order perturbation Feynman diagrams
for the phase contribution from one-gluon exchange between two
constituent quarks.}
\end{figure}
At the lowest order of the light-front time-order perturbation theory,
we have one gluon exchange contribution to the interaction kernel.
This can be expressed as a sum of all diagrams with gluon
connection between all possible pair of constituents in  the light-front
wave function. For example, the contribution from
the gluon exchange between the $i$th and $j$th quark
can be written as,
\begin{eqnarray}
K[k;\ell]_{ij}&=&\frac{\bar u_{\lambda_i}(x_i,k_{i\perp})}{\sqrt{x_i}}\gamma^\mu
\frac{u_{\lambda_i'}(y_i;\ell_{i\perp})}{\sqrt{y_i}}
d_{\mu\nu}\frac{\bar u_{\lambda_j}(x_j,k_{j\perp})}{\sqrt{x_j}}\gamma^\nu
\frac{u_{\lambda_j'}(y_j;\ell_{j\perp})}{\sqrt{y_i}} \nonumber\\
&&\times\left\{\frac{1}{P^--q^--k_i^--\ell_j^--\sum\limits_{\alpha\neq \{i, j\}}
k_\alpha^-+i\epsilon}
\frac{\theta(q^+)}{q^+}\right.\nonumber\\
&&\left.+\frac{1}{P^--q^{\prime-}-k_j^--\ell_i^--\sum\limits_{\alpha\neq \{i, j\}}
k_\alpha^-+i\epsilon}
\frac{\theta(q^{\prime+})}{q^{\prime+}}\right\} \ ,
\end{eqnarray}
where $\lambda$ represents the helicity for the associated quarks,
$q^+=k_j^+-\ell_j^+$ and $q^{\prime +}=k_i^+-\ell_i^+$,
and the color factors are implicit in the above
equation. Similar expression shall hold for the gluon
constituent in the wave function, and so the final results.
We  illustrate the  contribution
in the above calculations for $i=1$ and $j=2$. The first term
in the above bracket comes from Fig.~1(a), whereas the
second term comes from Fig.~1(b).
Moreover, at this particular order, quark number is conserved,
such that $n= n'$. The gluon polarization tensor is defined as,
\begin{equation}
d_{\mu\nu}=-g_{\mu\nu}+\frac{v_\mu \tilde q_\nu+v_\nu\tilde q_\mu}{[v\cdot q]} \ ,
\end{equation}
where $v$ is a light-like vector $v\cdot P=1$, $\tilde q$ differs from $q$ in the minus
component to take into account the instant propagator contribution~\cite{Lepage:1980fj}.
When one solves the Heisenberg eigenvalue problem for light-front
QCD as in discretized light-front quantization~\cite{Brodsky:1997de}, the light-cone
gauge singularity is regulated by the principal value prescription,
which corresponds to the antisymmetric boundary condition
for the gauge potential $A_\perp(x^-=+\infty)+A_\perp(x^-=-\infty)=0$.
However, this prescription will not result into an imaginary part for the
light-front wave function from the above
interaction kernel. In the following calculation,
we will choose the advanced boundary condition: $A_\perp(x^-=+\infty)=0$
whereas $A_\perp(x^-=-\infty)\neq 0$
in order to construct the augmented light-front wavefunction.
With this boundary condition, the
light-cone singularity will be regulated by~\cite{BelJiYua02},
\begin{equation}
\frac{v_\mu \tilde q_\nu+v_\mu\tilde q_\nu}{[v\cdot q]}|_{Adv.}=\frac{v_\mu\tilde q_\nu}{v\cdot q-i\epsilon} +\frac{v_\nu \tilde q_\mu}{v\cdot q+i\epsilon}\ ,
\end{equation}
where the momentum flow of $q$ is toward to the vertex $\nu$.
Clearly, this term contains a phase. The imaginary part is simple, and
proportional to a Delta function: $i\pi\delta(v\cdot q)$. Since we are
only interested in the imaginary part of the light-front wave function
amplitudes, we simply apply this Delta function to the interaction
kernel in Eq.~(3). In particular, we find that the dominant contribution comes
from the $d_{+\perp}$ components of the $d_{\mu\nu}$ tensor~\cite{BelJiYua02,mueller}.
All other contributions cancel out between
the above two terms or by themselves.
Another important consequence is that
the helicities are conserved in the interaction kernel:
$\delta_{\lambda_i\lambda_i'}\delta_{\lambda_j\lambda_j'}$.
After a little algebra, we obtain a rather simple result for the
imaginary part of the light-front wave function amplitude generated
from lowest order perturbation theory,
\begin{eqnarray}
{\cal I} \left[\psi_n(x_\alpha,k_{\alpha\perp})\right]=-\frac{\alpha_s}{2\pi}[{\rm C.F.}]
\int\frac{d^2q_\perp}{\vec{q}_\perp^2} \left(1-\frac{P^--\sum \ell^-}{P^--\sum k^-}\right)
\sum\limits_{i\neq j}\psi_n^{(ij)}(x_\beta;\ell_{\beta\perp}) \ ,
\end{eqnarray}
where $[{\rm C.F.}]$ represents the color-factor for the Feynman diagram in Fig.~1 and
$\psi_n^{(ij)}=\psi_n(x_\alpha;\ell_{i\perp}=k_{i\perp}-q_\perp,\ell_{j\perp}=
k_{j\perp}+q_\perp,\ell_{\beta \perp}|_{\beta\neq{i,j}}=k_{\beta\perp})$.
We emphasize that the wave function at the right
hand side only contains real part.

The above equation is
the main result of this paper.
It explicitly demonstrates that the light-front
wave function amplitudes contain an imaginary part if we choose
advanced boundary condition for the transverse component of the
gauge potential. If we choose the retarded boundary condition, we
obtain an opposite sign in the above equation.

\section{Applications to Pion and Nucleon}

The three-quark Fock state components have been
classified in Ref.~\cite{Ji:2002xn}. For these light-front
wave function amplitudes, we can apply the derivation
in the last section, and obtain the imaginary
part as,
\begin{eqnarray}
{\cal I}\left[\Psi_{qqq}(x_i,k_{i\perp})\right]&=&\frac{\alpha_s}{2\pi}{C_B}
\int {d^2\ell_{1\perp}d^2\ell_{2\perp}d^2\ell_{3\perp}}\delta^{(2)}
(\ell_{1\perp}+\ell_{2\perp}+\ell_{3\perp})\Psi_{qqq}(x_i,\ell_{i\perp})
\nonumber\\
&&~\times\left(1-
\frac{P^{-}-\sum \ell^-}{P^{-}-\sum k^{-}}\right)
\left[\frac{\delta^{(2)}(\ell_{3\perp}-k_{3\perp})}{(\vec{k}_{1\perp}-\vec{\ell}_{1\perp})^2}
+(2\leftrightarrow 3)+ (1\leftrightarrow 3)\right] \ ,
\end{eqnarray}
where $C_B={(N_c+1)}/{2N_c}$ and $\Psi$ represents the
general wave function amplitude constructed in Ref.~\cite{Ji:2002xn}.
By applying these results,
we are able to formulate the naive time-reversal-odd
quark distributions (such as the quark Sivers function)
in terms of the light-front wave function
amplitudes, by taking into account the above imaginary phase
using the above derivation~\cite{Ji:2002xn}.

For the quark-antiquark Fock component of pion, the result
can be further simplified as
\begin{eqnarray}
{\cal I}\left[\psi(x,k_\perp)\right]&=&\frac{\alpha_s}{2\pi}C_F\int\frac{d^2q_\perp}{\vec{q}_\perp^2}
\psi(x,k_\perp-q_\perp)  \left(1-\frac{x(1-x)M^2-(k_\perp-q_\perp)^2-m_q^2}{x(1-x)M^2-{k_\perp^2}-m_q^2}\right) ,
\end{eqnarray}
where we have chosen $P_\perp=0$ and assumed that the quark and antiquark
have the same mass $m_q$ and $C_F={(N_C^2-1)}/{2N_C}$.
In particular, if we are interested in the large transverse
momentum behavior of the light-front wave function amplitudes,
we can expand the interaction kernel in terms of $\ell_\perp/k_\perp$, and
keep the leading order contribution. By doing that,
we will obtain,
\begin{equation}
{\cal I}\left[\psi(x,k_\perp)\right]=\frac{\alpha_s}{2\pi}\frac{1}{\vec{k}_\perp^2}
C_F\phi(x)\ ,
\end{equation}
and $\phi(x)$ is the leading-twist distribution amplitude for Pion,
normalized by the leading Fock component light-front wave function
$\phi(x)=\int d^2\ell_\perp \psi(x,\ell_\perp) $.
Similar expressions can be found for the
quark-diquark model~\cite{BroHwaSch02}.

Although light-front wave functions
depend on the boundary condition of the gauge potential
in the light-cone gauge, physical observables cannot depend on
this choice because of gauge invariance~\cite{BelJiYua02,{Brodsky:2002ue}}.
In particular, the single-spin asymmetry in semi-inclusive deep
inelastic polarized proton deep inelastic scattering
$\ell p^\updownarrow \to \ell^\prime q X$
and the associated quark Sivers function can
be formulated simply as the overlap of  augmented LFWFs using the
advance boundary condition~\cite{Ji:2002xn}.
In particular, it is the phase difference between
the LFWFs for the $S$ and $P$-wave Fock components that
contributes to the quark Sivers function in the quark-diquark
model studied in Ref.~\cite{BroHwaSch02}. The
imaginary phases are calculated by using the general formalism
Eq.~(6) with similar expression as Eq.~(8).

The result for the Sivers single-spin asymmetry
using augmented LFWFs is identical to that found in Ref.~\cite{BroHwaSch02}
using conventional LFWFs (with the principal value boundary condition),
together with an explicit calculation of the final state phases which
arise from the rescattering of the struck quark with the spectator diquark
after the lepton-quark interaction.
This identity is possible since the final-state phase
due to rescattering is independent of the momentum
transferred in the lepton-quark interaction.
On the other hand,  If we choose the retarded boundary condition,
the augmented wave function will have opposite imaginary
part. However, under this boundary condition, we have to take
into account the final state interaction effects (the gauge link
contributions from the quark distributions), but again, this leads to
the same result compared to that using the advanced boundary condition.

Similar conclusions hold for the small-$x$ parton
distribution calculated in~\cite{Brodsky:2002ue}.   We leave this topic
for a future publication.

\section{Summary and Discussions}


We  have use  light-front time-ordered perturbation theory to obtain
augmented light-front wave functions  which contain an imaginary
phase which depends on the choice of advanced or retarded
boundary condition for the gauge potential in light-cone gauge.
We have applied these results to construct augmented wavefunctions for the three-quark or quark-diquark
Fock state components of nucleon
and  the quark-antiquark component of the pion.
We obtain the leading-twist quark Sivers function from these augmented light-front wavefunctions,
by applying the overlap formalism~\cite{Ji:2002xn}.
The result is identical to the
explicit calculation~\cite{BroHwaSch02} of the single spin-asymmetry in semi-inclusive deep inelastic lepton-polarized nucleon scattering where the required phases come from the final-state rescattering of the struck quark  with the nucleon spectators.


This work was supported in part by the U.S. Department of Energy
under contracts DE-AC02-05CH11231 and DE-AC02-76SF00515, and by
the Research Infrastructure Integrating Activity
``Study of Strongly Interacting Matter'' (acronym HadronPhysics2, Grant
Agreement n. 227431) under the Seventh Framework Programme of the
European Community. We are grateful to RIKEN,
Brookhaven National Laboratory and the U.S. Department of Energy
(contract number DE-AC02-98CH10886) for providing the facilities
essential for the completion of this work. SLAC-PUB-13874

\end{document}